\documentclass[pra,twocolumn,showpacs,preprintnumbers,letterpaper]{revtex4}
\usepackage{setspace}
\usepackage{graphicx}
\usepackage{amsmath}
\usepackage{epsfig}
\usepackage{amssymb}

\begin{document}

%
%
%
%
%
\def\bra#1{\mathinner{\langle{#1}|}}
\def\ket#1{\mathinner{|{#1}\rangle}}
\def\braket#1{\mathinner{\langle{#1}\rangle}}
\def\Bra#1{\left<#1\right|}
\def\Ket#1{\left|#1\right>}
{\catcode`\|=\active
  \gdef\Braket#1{\left<\mathcode`\|"8000\let|\BraVert {#1}\right>}}
\def\BraVert{\egroup\,\mid@vertical\,\bgroup}
%

{\catcode`\|=\active
  \gdef\set#1{\mathinner{\lbrace\,{\mathcode`\|"8000\let|\midvert #1}\,\rbrace}}
  \gdef\Set#1{\left\{\:{\mathcode`\|"8000\let|\SetVert #1}\:\right\}}}
\def\midvert{\egroup\mid\bgroup}
\def\SetVert{\egroup\;\mid@vertical\;\bgroup}


\title{Correlated Gaussian Hyperspherical Method for Few-Body Systems}
\author{Javier von Stecher}
\author{Chris H. Greene}
\affiliation{JILA and Department of Physics, University of Colorado, Boulder, Colorado
80309-0440}

\begin{abstract}
We develop an innovative numerical technique to describe few-body systems.
Correlated Gaussian basis functions are used to expand the channel functions
in the hyperspherical representation. The method is proven to be robust and
efficient compared to other numerical techniques. The method is applied to
few-body systems with short range interactions, including several examples
for three- and four-body systems. Specifically, for the two-component,
four-fermion system, we extract the coefficients that characterize its
behavior at unitarity.
\end{abstract}

\maketitle

\section{Introduction}

Ultracold gases in traps or optical lattices have opened new possibilities
in the study of strongly correlated quantum systems. From the rich few-body
physics of the Efimov effect~\cite%
{braaten2006ufb,efim70,esry1999rta,kraemer2006eeq} to the fascinating
many-body physics of the BCS-BEC crossover~\cite%
{eagl69,legg80,nozieres1985bca,greiner2003emb,jochim2003bec,
zwierlein2003obe,bour04,regal2004orc,zwierlein2004cpf,zwierlein2005vas}%
, experimentalists are now able to realize a wide variety of physical
systems of great interest for the atomic, nuclear, and condensed matter
communities. In particular, the pureness and controllability of cold atoms
in optical lattices~\cite{jaksch1998cba,paredes2004fa,kinoshita2004ood} make
them perfect candidates for the experimental implementation of condensed
matter models [see Ref.~\cite{bloch2008mbp} and references therein]. In all
these systems, the rich physics that governs a few interacting atoms is
crucial for understanding recent experiments.

For that reason, extensive efforts have concentrated on the development of
an accurate description of few-body systems. Encouraging advances have been
achieved in the last decade in the understanding ultracold three-body
problem~\cite%
{braaten2006ufb,esry1999rta,PhysRevLett.93.143201,PhysRevA.67.010703}. These
studies have demonstrated the importance of three-body recombination and
relaxation processes and have determined the effective interaction in
atom-dimer collisions. Some of these techniques were subsequently extended
to four-body systems~~\cite{petrov2004wbd,deltuva2007fbc,stech07,dincao08,wang2008etf}, in a few applications.
However, the physics of the four-body problem that are far richer and more
complicated. Also, it is a very challenging numerical problem and for that
reason it has remained largely unsolved except in very limited regimes.
Here, we present a novel numerical method to handle few-body systems that
can be used to efficiently describe four-body systems, through a combination
of different techniques.

Even though several techniques have been developed in recent decades to
provide solutions for few-body systems ~\cite%
{faddeev1960mat,suzuki1998sva,malfliet1969sfe,yaku67,mace68}, not many of
them have been applied to numerically solve the Schr\"odinger equation for
systems with more than three particles. Among these methods, the correlated
Gaussian (CG) technique~\cite%
{singer1960uge,boys1960ifv,kukulin1977svm,varga1996svm,varga1995psf,varga1997sfb,varga1994mmd}
in particular has proven to be capable of describing a trapped few-body
system with short-range interactions.
Because of the simplicity of the matrix element calculation, the CG method
provides an accurate description of the ground and excited states up to $N=6$
particles~\cite{varga1995psf, varga1997sfb,blumePRL07}.
However, the CG method as previously implemented can only describe bound
states. For this reason, previous studies have focused on trapped systems
where all the eigenstates are discrete~\cite%
{vonstech07,vonstech08a,stech07,blumePRL07}. In fact, the CG method requires
a nontrivial extension in order to describe the continuum and the rich
behavior of atomic collisions, such as dissociation, rearrangement, and
recombination processes.

The hyperspherical representation, in fact, provides an appropriate
framework that can treat the continuum~\cite{Delves59,Delves60,mace68,
Lin95, Shitikova77, Chapuisat92}. In the adiabatic hyperspherical
representation, the Hamiltonian is diagonalized as a function of the
hyperradius $R$, reducing the Schr\"odinger equation to a set of coupled
equations in a single variable, with a series of different effective
potentials and couplings. The asymptotic behavior of the channel potentials
describes different dissociation or fragmentation pathways and provides a
suitable framework for analyzing collision physics. These solutions can be
readily combined with scattering methods such as the R-matrix approach~\cite%
{aymar1996mrs,zatsarinny2004bsb,sunderland2002prm}
to provide an accurate description of the collisional dynamics. However, the
standard hyperspherical methods expand the hyperangular channel functions in
a B-spline or finite element basis set~\cite%
{pack1987qrs,zhou1993hac,esry1996ahs,suno2002tbr}, and the calculations
become very computationally demanding for $N>3$ systems.

It is therefore natural to combine the scalability of the CG method with the
advantages of the hyperspherical representation. In this article, we present
an innovative way to achieve this combination, in what we term the
correlated Gaussian hyperspherical method (CGHS). This method uses CG basis
functions to expand the channel functions in the hyperspherical
representation. We show that also in this case, the matrix element
evaluation is greatly simplified thanks to the simple form of the CG basis
functions. Furthermore, thanks to the explicit correlation incorporated in
these basis functions, only a relatively small basis set is needed to
achieve convergence of the lowest channel functions even in the strongly
interacting regime.

To illustrate the power of the CGHS method, we carry out calculations for $%
N=3,4$-particle systems in the strongly interacting regime. First, we
analyze systems of three-bosons or three fermions at unitarity, and show
that the method recovers results that agree with semi-analytical
predictions. Then, we consider the two-component four-fermion system, in the
large and positive scattering length regime, and reproduce the lowest
potential curves from Ref.~\cite{dincao08}. The CGHS provides a larger
number of channels which would allow the calculation of scattering events
not considered in Ref.~\cite{dincao08}. Finally, we focus on the universal
behavior of four-fermions at unitarity. In this regime, the energies of the
trapped system are trivially determined by the hyperspherical potential
curves~\cite{wern06,blumePRL07}. Therefore, we can compare our calculations
with predictions for the trapped system~\cite{stech07,
chan07,vonstechtbp,alhassid08,StetcuCM07}. Our results improve and extend
these previous predictions, and characterize the 20 lowest potential curves
for even parity and vanishing orbital angular momentum.

This article continues as follows. First, we review both the CG and
hyperspherical methods in Sec.~\ref{theoback}. In Sec.~\ref{CGHSSec}, we
introduce the main idea of the CGHS method, leaving some details of the
implementation for the Appendix~\ref{App}. Sec.~\ref{Results} presents our
results for three-body systems and for the four-fermion system. Finally,
Sec.~\ref{conclusions} presents our conclusions.

\section{Theoretical background}

\label{theoback}

This section discusses the general problem to be solved and reviews the
correlated Gaussian method. Subsec.~\ref{hyperRepSec} presents the general
formalism of the hyperspherical representation and describes how to
numerically solve the Schr\"odinger equation in this representation using a
correlated Gaussian basis set expansion.

The methods described in this article solve the time-independent
Schr\"odinger equation for a Hamiltonian of the form
\begin{eqnarray}  \label{eq_ham}
H = \sum_{i} \left(\frac{-\hbar^2}{2m_i} \nabla_i^2 +V_{ext}(\mathbf{r}_i)
\right) + \sum_{i,j} V_0(r_{ij}).
\end{eqnarray}
where $V_{ext}$ is an external trapping potential and $V_0$ is the
interaction potential. The form of the Hamiltonian can be varied depending
on the particular problem we are considering. In the CG method one will
usually consider a spherically-symmetric harmonic trapping potential $%
V_{ext}(\mathbf{r})=\frac{1}{2} m_i \omega^2 \mathbf{r}_i^2 $ but in
hyperspherical calculations we usually consider a free system ($V_{ext}(%
\mathbf{r})=0$). We can always include the harmonic trapping potential in
the final step of the hyperspherical calculation, since it is a purely
hyperradial potential. Depending on the particular problem considered, the
interacting particles will change.
For example, all particles interact with each other in identical boson
systems but only opposite-spin fermions interact in two-component Fermi
systems (except in a few problems involving $p$-wave Fano-Feshbach
resonances). Also, in many cases, the center-of-mass motion decouples from
the more interesting internal degrees of freedom, and it is preferable to
use a set of Jacobi coordinates rather than the usual single-particle
coordinates. All such options can be treated using the method presented
below.

\subsection{Correlated Gaussian Method}

\label{CGSec}

Different types of Gaussian basis functions have long been used in many
different areas of physics. In particular, the usage of Gaussian basis
functions is one of the key elements of the success of \textit{ab initio}
calculations in quantum chemistry. The idea of using an explicitly
correlated Gaussian to solve quantum chemistry problems was introduced in
1960 by Boys~\cite{boys1960ifv} and Singer~\cite{singer1960uge}. The
combination of a Gaussian basis and the stochastical variational method
(SVM) was first introduced by Kukulin and Krasnopol'sky~\cite{kukulin1977svm}
in nuclear physics and was extensively used by Suzuki and Varga~\cite%
{varga1996svm,varga1995psf,varga1997sfb,varga1994mmd}. These methods were
also used to treat ultracold many-body Bose systems by Sorensen, Fedorov and
Jensen~\cite{sorensen2005cgm}. A detailed discussion of both the SVM and CG
methods can be found in a thesis of Sorensen \cite{sorensen2005cmb} and, in
particular, in the book by Suzuki and Varga~\cite{suzuki1998sva}. In the
following, we highlight the main ideas of the CG method.

Consider a set of coordinate vectors that describe the system $\{\mathbf{x}%
_1,...,\mathbf{x}_N\}$. In this method, the eigenstates are expanded in a
set of basis functions,
\begin{equation}
\Psi (\mathbf{x}_1,\cdots,\mathbf{x}_N) = \sum_A C_A\,\Phi _A(\mathbf{x}%
_1,\cdots,\mathbf{x}_N).  \label{TotalWF}
\end{equation}
Here $A$ specifies a matrix with a particular set of parameters that
characterize the basis function. It is convenient to introduce the following
ket notation, $\Phi _A(\mathbf{x}_1,\cdots,\mathbf{x}_N)=\braket{%
\mathbf{x}_1,\cdots,\mathbf{x}_N|A}$. Solution of the time-independent
Schr\"odinger equation in this basis set reduces the problem to one of
diagonalizing the Hamiltonian matrix:
\begin{equation}
\mathcal{H}\vec{C}_i =E_i \mathcal{O}\vec{C}_i  \label{HamMatrix}
\end{equation}
Here, $E_i$ are the energies of the eigenstates, $\vec{C}_i$ is a vector
formed with the coefficients $C_A$ and $\mathcal{H}$ and $\mathcal{O}$ are
matrices whose elements are $\mathcal{H}_{BA}=\braket{B|\mathcal{H}|A}$ and $%
\mathcal{O}_{BA}=\braket{B|A}$. For a 3D system, the evaluation of these
matrix elements involves $3N$-dimensional integrations which are in general
very expensive to compute. Therefore, the effectiveness of the basis set
expansion method relies mainly on the appropriate selection of the basis
functions. As we will see, the CG basis functions permit fast evaluation of
overlap and Hamiltonian matrix elements, and they are flexible enough to
correctly describe physical states.

To reduce the dimensionality of the problem we can take advantage of its
symmetry properties. Since the interactions considered are spherically
symmetric, the total angular momentum, $L$, is a good quantum number. For
simplicity, we will restrict ourselves to $L=0$ solutions. This restriction
allows us to reduce the Hilbert space by introducing restrictions on the
basis functions. In particular, if the basis functions only depend on the
interparticle distances, then Eq.~(\ref{TotalWF}) can only describe states
with zero angular momentum and positive parity ($L^P=0^+$).
Furthermore, we can recognize that the center-of-mass motion decouples from
the system. In such cases, the CG basis functions take the form
\begin{equation}
\Phi _{A}(\mathbf{x}_1,\cdots,\mathbf{x}_N) =\psi_0(\mathbf{R}_{CM})\mathcal{%
S}\left\{\exp\left(-\sum_{j>i=1}^N \alpha_{ij} r_{ij}^2/2\right)\right\},
\label{BF}
\end{equation}
where $\mathcal{S}$ is a symmetrization operator and $r_{ij}$ is the
interparticle distance between particles $i$ and $j$.
Here, $\psi_0$ is the ground state of the center-of-mass motion. For trapped
systems, $\psi_0$ takes the form, $\psi_0(\mathbf{R}%
_{CM})=e^{-R_{CM}^2/2(a_{ho}^{M})^2}$. Because of its simple Gaussian form, $%
\psi_0$ can be absorbed in the exponential factor. Thus, in a more general
way, the basis function can be written in terms of a matrix $A$ that
characterizes them,
\begin{multline}  \label{Basis1}
\Phi_A(\mathbf{x}_{1},\mathbf{x}_{2},...,\mathbf{x}_{N})= \mathcal{S}%
\left\{\exp(-\frac{1}{2} \mathbf{x}^T.A.\mathbf{x})\right\} \\
=\mathcal{S}\left\{\exp(-\frac{1}{2} \sum_{j,i=1}^N A_{ij} \mathbf{x}_i.%
\mathbf{x}_j)\right\},
\end{multline}
where $\mathbf{x}=\{\mathbf{x}_1,\mathbf{x}_2,...,\mathbf{x}_N\}$, and $A$
is a symmetric matrix. The matrix elements $A_{ij}=A_{ji}$ can be expressed
in terms of the $\alpha_{ij}$.
Because of the simplicity of the basis functions, Eq.~(\ref{BF}), the matrix
elements of the Hamiltonian can be calculated analytically.

The analytical evaluation of the matrix elements is enabled by selecting the
set of coordinates that simplifies the evaluations. For basis functions of
the form of Eq.~(\ref{Basis1}), the matrix elements are characterized by a
matrix $M$ in the exponential. Then the matrix element integrand greatly
simplifies if we rewrite it in terms of the coordinate vectors that
diagonalize that matrix $M$. This change of coordinates permits, in many
cases, the analytical evaluation of the matrix elements. The explicit
evaluation of several matrix elements can be found in Refs.~\cite{sorensen2005cmb,suzuki1998sva}.

Two properties of the CG method deserve mention at this point. First, the CG
method does not rely on any approximation other than basis set truncation,
and the solutions can be systematically improved. The accuracy of the
results are only limited by numerical issues related to linear dependence of
the basis set. Secondly, the basis functions $\Phi_A$ are square-integrable
only if the matrix $A$ is positive definite. This ensures that the wave function decays in all degrees of freedom. We can
further restrict the basis functions by introducing real widths $d_{ij}$
such that $\alpha_{ij}=1/d_{ij}^2$. With this transformation, we ensure that
$A$ is positive definite. Furthermore, each such width is proportional to
the mean interparticle distances covered by that basis function. Thus, it is
relatively easy to select the widths after considering the physical length
scales relevant to the problem. Even though we have restricted the Hilbert
space with this transformation, we have numerical evidence that the results
converge to the exact eigenvalues.

The linear dependence in the basis set causes problems in the numerical
diagonalization of the Hamiltonian matrix, Eq.~(\ref{HamMatrix}). To
minimize these linear dependence problems we restrict the basis function so
that the overlap between any two normalized basis functions is below some
cutoff value. The other method we use to eliminate linear dependence applies
a linear transformation to produce a smaller orthonormal basis set.

Finally, we stress the importance of making an appropriate selection of the
interaction potential. For the problems considered in this article, the
interactions are expected to be characterized only by the scattering length,
i.e., to be independent of the shape of the potential. For that reason, we
can select a model potential that permits rapid evaluation of the matrix
elements. We have found that a model potential with a Gaussian form,
\begin{eqnarray}  \label{potNumChap}
V_0(r)=-V_0\exp \left(- \frac{r^2}{2r_0^2} \right),
\end{eqnarray}
is particularly suitable for this basis set expansion since it can be
absorbed in the exponential form of the wave functions for matrix element
evaluation. If the range $r_0$ is much smaller than the scattering length,
then the interactions are effectively characterized only by the scattering
length. The scattering length is tuned by changing the strength of the
interaction potential, $V_0$, while the range, $r_0$, of the interaction
potential remains unchanged. This is particularly convenient in this method
since it implies that we only need to evaluate the matrix elements once and
we can use them to solve the Schr\"odinger equation at any given potential
strength (or scattering length). Of course, this procedure will give
accurate results only if the basis set is sufficiently flexible and complete
to describe the different configurations that appear at different scattering
lengths.

In general, this method includes five basic steps: generation of the basis
set, evaluation of the matrix elements, elimination of linear dependence,
evaluation of the eigenvalue spectrum, followed by a study of stability and
convergence. The stochastic variational method (SVM) Refs.~\cite%
{sorensen2005cmb,suzuki1998sva} combines the first three of these steps in
an optimization procedure where the basis functions are selected randomly.

\subsection{Hyperspherical representation}

\label{hyperRepSec}

The main objective of the hyperspherical method is to solve the
time-independent Schr\"odinger equation in a convenient and efficient way
that also provides insight into the relevant reaction pathways by which
various collision processes can occur. the first step involves calculation
of eigenvalues and eigenfunctions of the fixed-hyperradius Hamiltonian,
which defines the adiabatic hyperspherical representation. These eigenvalues
and eigenfunctions are then used to construct a set of one-dimensional
coupled equations in the hyperradius $R$. The hyperradius is a collective
coordinate related to the total moment of inertia of the system\cite%
{Shitikova77, fano1981}. In a system described by $N$ coordinate vectors $%
\mathbf{r}_1,\ldots,\mathbf{r}_N$, the hyperradius $R$ is defined by
\begin{equation}  \label{HR}
\mu R^2=\sum_{i=1}^N m_i \mathbf{r}_i^2.
\end{equation}
Here, $\mu$ is an arbitrary mass factor called the hyperradial reduced mass~\cite{fnote1}%
and $m_i$ are the masses corresponding to the particle $i$. The remaining
coordinates are described by a set of hyperangles, collectively denoted $%
\Omega$.

The total number of spatial dimensions of this $N$-particle system is $%
d=3\,N $. The total wave function $\psi$ is rescaled by $R$, $%
\Psi=R^{(d-1)/2}\psi$, so that the hyperradial equation resembles a coupled
one-dimensional Schrodinger equation. In the adiabatic representation, the
wave function $\Psi_E(R,\Omega)$ is expanded in terms of a complete
orthonormal set of angular wave functions $\Phi_\nu$ and radial wave
functions $F_{\nu E}$, such that
\begin{equation}  \label{ExpHR}
\Psi_E(R,\Omega)=\sum_{\nu}F_{\nu E}(R)\Phi_\nu(R;\Omega).
\end{equation}
The adiabatic eigenfunctions, or channel functions $\Phi_\nu$, depend
parametrically on $R$ and are eigenfunctions of a $3\,N-1$ partial
differential equation (which reduces to $3N-4$ dimensions if the
center-of-mass motion is removed explicitly):
\begin{multline}  \label{ChanHR}
\left(\frac{\hbar^2\Lambda^2}{2\mu R^2}+\frac{(d-1)(d-3)\hbar^2}{8\mu R^2}%
+V(R,\Omega)\right)\Phi_\nu(R;\Omega) \\
=U_\nu (R)\Phi_\nu(R;\Omega).
\end{multline}
%
Here, $\Lambda$ is the grand angular momentum operator, which is related to
the kinetic term by
\begin{equation}  \label{kinHR}
-\sum_i\frac{\hbar^2\nabla_i^2}{2 m_i}=-\frac{\hbar^2}{2\mu}\frac{1}{R^{d-1}}
\frac{\partial}{ \partial R} R^{d-1}\frac{\partial}{ \partial R}+\frac{%
\Lambda^2\hbar^2}{2\mu R^2}.
\end{equation}

The $U_\nu(R)$ obtained in Eq.~(\ref{ChanHR}) are effective hyperradial
potential curves that appear in a set of coupled one-dimensional
differential equations:
\begin{multline}  \label{HRradial}
\left[-\frac{\hbar^2}{2\mu}\frac{d^2}{dR^2}+U_\nu(R)\right]F_{\nu E}(R) \\
-\frac{\hbar^2}{2\mu}\sum_{\nu^{\prime}}\left[2P_{\nu\nu^{\prime}}(R)\frac{d%
}{dR} +Q_{\nu\nu^{\prime}}(R)\right]F_{\nu^{\prime}E}(R)=EF_{\nu E}(R).
\end{multline}
These differential equations [Eq.~(\ref{HRradial})] are coupled through the $%
P_{\nu\nu^{\prime}}(R)$ and $Q_{\nu\nu^{\prime}}(R)$ couplings defined as
\begin{gather}  \label{couplings}
P_{\nu\nu^{\prime}}(R)=\braket{\Phi_\nu(R;\Omega)|\frac{\partial}{\partial
R}|\Phi_{\nu'}(R;\Omega)}\Big|_R, \\
Q_{\nu\nu^{\prime}}(R)=\braket{\Phi_\nu(R;\Omega)|\frac{\partial^2}{\partial
R^2}|\Phi_{\nu'}(R;\Omega)}\Big|_R.
\end{gather}

Since the basis set expansion of the wave function, Eq.~(\ref{ExpHR}), is
complete in the $3\,N$-dimensional space, Eqs.~(\ref{ChanHR}) and (\ref%
{HRradial}) reproduce exactly the original $d$-dimensional Schr\"odinger
equation. As in most numerical methods, the solutions are approximated by
truncating the Hilbert space. In this case, the Hilbert space is truncated
by considering a finite number of channels in Eq.~(\ref{HRradial}). This
approximation is easily tested by analyzing convergence with respect to the
number of channels included in the calculation.

The utility of the hyperspherical representation relies on the assumption
that the wavefunction variation with the hyperradius $R$ is smooth. In such
cases, only a few channels are relevant, and the couplings are small and
vary smoothly with $R$. Furthermore, a fairly good approximation to the
solutions can be achieved by truncating the expansion in Eq.~(\ref{ExpHR})
to a single term:
\begin{equation}  \label{adiaWF}
\Psi_E(R,\Omega)=F_{\nu E}(R)\Phi_\nu(R;\Omega).
\end{equation}
This adiabatic hyperspherical approximation leads to an effective
one-dimensional Schr\"odinger equation,
\begin{equation}  \label{HRadia}
\left[-\frac{\hbar^2}{2\mu}\frac{d^2}{dR^2}+W_\nu(R)\right]F_{\nu
E}(R)=EF_{\nu E}(R),
\end{equation}
where the effective potential is
\begin{equation}  \label{WHRradial}
W_\nu(R)=U_\nu(R)-\frac{\hbar^2}{2\mu}Q_{\nu\nu}(R).
\end{equation}
Here, the first term is the hyperradial potential curve, and the second term
is ``adiabatic correction'', i.e. the repulsive kinetic contribution of the
hyperradial dependence of the channel function. If the potential curves are
well-separated and have no strong avoided crossings in the relevant range of
energy and radius, then the adiabatic approximation can be quite accurate
for the lower states in any given potential curve. This approximation comes
from a truncation of the Hilbert space and, for that reason, obeys the
variational principle. Any discrete energy eigenvalue obtained with this
method is an upper bound of the exact energy level, in the sense of the
Hylleraas-Undheim theorem. An approximate description of the spectrum can be achieved by combining the
energies obtained from the adiabatic approximation applied to each channel
separately. For example, bound states of excited potential curves which are
above the lowest fragmentation threshold would represent quasi-bound states.
This approach is equivalent to neglecting all off-diagonal couplings in Eq.~(%
\ref{HRradial}) and produces an approximate spectrum which is not
variational.
Another useful approximation is obtained by neglecting the second term in
Eq.~(\ref{WHRradial}), i.e., replacing $W_0(R)$ by $U_0(R)$ in Eq.~(\ref%
{HRadia}). This is usually called the hyperspherical Born-Oppenheimer
approximation. As in the standard Born-Oppenheimer approximation for a
diatomic molecule, the approximate energy obtained in this manner represents
a lower bound to the exact ground state energy~\cite{Coelho91}.

Next, we show how Eq.~(\ref{ChanHR}) is solved, and how the $%
P_{\nu\nu^{\prime}}$ and $Q_{\nu\nu^{\prime}}$ are evaluated.

\subsection{Expansion of the channel function in a basis set}

\label{Exp}

In the hyperspherical method (see Sec.~\ref{hyperRepSec}), channel functions
are eigenfunctions of the adiabatic Hamiltonian $\mathcal{H}_A(R;\Omega)$,%
\begin{equation}  \label{LamHRnew}
\mathcal{H}_A(R;\Omega)\Phi_\nu(R;\Omega)=U_\nu(R)\Phi_\nu(R;\Omega).
\end{equation}
The eigenvalues of this equation are the hyperspherical potential curves $%
U_\nu (R)$, which serve as readily visualizable reaction pathways. The
adiabatic Hamiltonian has the form:
\begin{equation}
\mathcal{H}_A(R;\Omega)=\frac{\hbar^2\Lambda^2}{2\mu R^2}+\frac{%
(d-1)(d-3)\hbar^2}{8\mu R^2}+V(R,\Omega).
\end{equation}
Here, $d=3N_J$ where $N_J$ is the number of Jacobi coordinate vectors.

A standard way to solve Eq.~(\ref{LamHRnew}) is to expand the channel
functions in a basis,
\begin{equation}
\ket{\Phi_\mu(R;\Omega)}=\sum_{i}\ket{B_i(R;\Omega)}c_{i\mu }(R).
\label{Phiexp}
\end{equation}%
Here $\mu $ labels the channel function. The $\ket{B_i(R;\Omega)}$ are the
basis functions. With this expansion, Eq.~(\ref{LamHRnew}) reduces to the
eigenvalue equation
\begin{equation}
\mathcal{H}_{A}(R)\vec{c}_{\mu }=U_{\mu }(R)\mathcal{O}(R)\vec{c}_{\mu }.
\label{HRa}
\end{equation}%
The $\mu $-th column vector $\vec{c}_{\mu }=\{c_{i\mu }\},i=1,...D$, where $%
D $ is the dimension of the basis set. $\mathcal{H}_{A}$ and $\mathcal{O}$
are the Hamiltonian and overlap matrices whose matrix elements are given by
\begin{gather}
\label{Hrr}\mathcal{H}_A(R)_{ij}=\braket{B_i|\mathcal{H}_A(R;\Omega)|B_j}\Big|_R,\\
\label{Orr}\mathcal{O}(R)_{ij}=\braket{B_i|B_j}\Big|_R.
\end{gather}

Once the hyperradial potential curves are calculated, we still need to
evaluate the $P$ and $Q$ non-adiabatic couplings between the channel functions (Eq.~(\ref{couplings}) in Sec.~\ref{hyperRepSec}). To evaluate the $Q$ coupling, we use the identity
\begin{equation}
Q_{\nu \mu }(R)=-\tilde{Q}_{\nu \mu }(R)+\frac{\partial P_{\nu \mu }(R)}{\partial R},
\end{equation}
where
\begin{equation}
\tilde{Q}_{\nu \mu }=\braket{\frac{\partial}{\partial R}\Phi_\nu(R)|\frac{\partial }{\partial R} \Phi_\mu(R)}.
\end{equation}%
 Thus, we can obtain all the couplings from the evaluation of $P$ and $\tilde{Q}$.
In the basis set expansion, $P$ and $\tilde{Q}$ can be calculated
using matrix multiplication. With the expansion in Eq.~(\ref{Phiexp}),
\begin{equation}
\ket{\Dot{\Phi}_\mu(R)}=\sum_{i}\ket{B_i}\Dot{c}_{i\mu }+\ket{\Dot{B}_i}%
c_{i\mu }.  \label{PhiExpder}
\end{equation}%
Here and in the following, we have omitted the radial and angular
dependence of functions, and we have introduced the notation $\Dot{F}$ for
the derivative of $F$ with respect to $R$. The $P$ coupling takes the form
\begin{equation}
P_{\nu \mu }=\sum_{ij}c_{\nu j}^{T}\braket{B_j|B_i}\Dot{c}_{i\mu }+c_{\nu
j}^{T}\braket{B_j|\Dot{B}_i}c_{i\mu }=\vec{c}_{\nu }^{T}\mathcal{O}\Dot{\vec{%
c}}_{\mu }+\vec{c}_{\nu }^{T}\mathcal{P}\vec{c}_{\mu }.  \label{P1}
\end{equation}%
where $\mathcal{P}(R)$ is defined later in Eq.~(\ref{PQrr}). The same procedure can
be done for the $\tilde{Q}$ matrix elements with
\begin{multline}
\tilde{Q}_{\nu \mu }=\sum_{ij}\Dot{c}_{\nu j}^{T}\braket{B_j|B_i}\Dot{c}_{i\mu }+%
\Dot{c}_{\nu j}^{T}\braket{B_j|\Dot{B}_i}c_{i\mu }  \label{Q1} \\
+c_{\nu j}^{T}\braket{\Dot{B}_j|B_i}\Dot{c}_{i\mu }+c_{\nu j}^{T}%
\braket{\Dot{ B}_j|\Dot{B}_i}c_{i\mu }
\end{multline}%
and can also be written in terms of matrix multiplications:
\begin{equation}
\tilde{Q}_{\nu \mu }=\Dot{\vec{c}}_{\nu }^{T}\mathcal{O}(R)\Dot{\vec{c}}_{\mu }+\Dot{%
\vec{c}}_{\nu }^{T}\mathcal{P}(R)\vec{c}_{\mu }+\vec{c}_{\nu }^{T}\mathcal{P}%
^{T}(R)\Dot{\vec{c}}_{\mu }+\vec{c}_{\nu }^{T}\tilde{\mathcal{Q}}(R)\vec{c}_{\mu }.
\label{Q2}
\end{equation}%
In Eqs.~(\ref{P1}, \ref{Q2}) we have used the overlap matrix $\mathcal{O}$
and defined the matrices $\mathcal{P}$ and $\tilde{\mathcal{Q}}$ whose matrix
elements are
\begin{equation}
\mathcal{P}(R)_{ij}=\braket{B_i(R)|\Dot{ B}_j}\,\,\mbox{and}\,\,\,\tilde{\mathcal{Q}}%
(R)_{ij}=\braket{\Dot{ B}_i|\Dot{ B}_j}.  \label{PQrr}
\end{equation}

The derivatives of the $\Dot{c}_{i\mu }(R)$ coefficients that form the $\Dot{%
\vec{c}}_{\mu }$ are calculated numerically using the three point rule. 

\section{Correlated Gaussian Hyperspherical method}

\label{CGHSSec}

As we have seen in the previous section, the implementation of
hyperspherical calculations requires the evaluation of the Hamiltonian
matrix elements at fixed $R$ (Eqs.~\ref{Hrr} and \ref{Orr}). This is one of
most time consuming part of the calculation which for an $N=4$ system
requires a 5 dimensional integration. Thus, we need to find an efficient way
to evaluate Hamiltonian matrix elements at fixed $R$. As a prelude, we first
review how multidimensional matrix elements evaluations reduce to analytical
forms in the standard CG method. This will be the key to evaluating matrix
elements in the hyperspherical variant of this method.

In the CG method, we select, for each matrix element evaluation, a set of
coordinate vectors that simplifies the integration, i.e., the set of
coordinate vectors that diagonalize the basis matrix $M$ which characterizes
the matrix element. The flexibility to choose the best set of coordinate
vectors for each matrix element evaluation is crucial for the economy of the
CG method.

This selection of the optimal set of coordinate vectors is formally applied
by an orthogonal transformation from an initial set of vectors $\mathbf{x}=\{%
\mathbf{x}_1,...,\mathbf{x}_N\}$ to a final set of vectors $\mathbf{y}=\{%
\mathbf{y}_1,...,\mathbf{y}_N\}$: $T\mathbf{x}=\mathbf{y}$, where $T$ is the
orthogonal transformation matrix. The hyperspherical method is particularly
suitable for such orthogonal transformations because the hyperradius $R$ is
an invariant under them. Consider the hyperradius defined in terms of a set
of mass-scaled Jacobi vectors~\cite%
{Delves59,Delves60,suno2002tbr,mehta2007haf}, $\mathbf{x}=\{\mathbf{x}_1,...,%
\mathbf{x}_N\}$,
\begin{equation}
\mu R^2=\mu\sum_i\mathbf{x_i}^2,
\end{equation}
If we applied an orthogonal transformation to a new set of vectors $\mathbf{y%
}$, then
\begin{equation}
\mu R^2=\mu\sum_i\mathbf{x_i}^2=\mu\mathbf{y}T^TT\mathbf{y}=\mu\sum_i\mathbf{%
y_i}^2
\end{equation}
where we have used the fact that $T^TT=I$, and $I$ is the identity.
Therefore, in the hyperspherical framework we can also select the most
convenient set of coordinate vectors for each matrix element evaluation.
This will be the key to reducing the dimensionality of the matrix element
integration. This transformation amounts to selecting, for each matrix
element evaluation, the set of hyperangles ($\Omega$) that simplifies the
matrix-element evaluation.

As an example of how the dimensionality of matrix-element integration is
thereby reduced, consider an $L=0$ three-dimensional $N$-particle system
with the center of mass removed. It can be shown that this technique reduces
a $(3N-7)$ numerical integration~\cite{fnote2}
to a sum over the symmetrization permutation of ($N-3$) numerical integrations. This result
implies that for $N=3$ the matrix element evaluation can be done
analytically (see Appendix~\ref{unsym3}) and that for $N=4$, it requires a
sum of one-dimensional numerical integrations~\cite{vonstechThesis}.

Once the basic idea of the appropriate change of variables for each matrix
element calculation is understood, the actual calculation of the matrix
elements using correlated Gaussian basis function is straightforward.
Appendix~\ref{unsym3} shows, as an example, how the matrix elements can be
calculated analytically for a three particle system (the calculation of the
matrix elements for $N=4$ are not presented here but can be found in Ref.~%
\cite{vonstechThesis}). Finally, Appendix~\ref{GenCon} discuss in general
how this method is implemented.

\section{Results}

\label{Results} In this section, we present CGHS results for $N=3,4$. First,
we analyze two different $N=3$ systems and compare them with analytical
predictions. Then, we present four-fermion potential curves and compare them
with recent predictions~\cite{dincao08}. Finally, we characterize the
four-fermion $L=0$ potential curves at unitarity and extract the $s_\nu$
coefficient that characterize the universal regime.

To test the CGHS method, we calculate the hyperspherical potential curves at
unitarity for three interacting bosons. For zero-range interactions, the
potential curves at unitarity are inversely proportional to the hyperradius.
For example, the lowest potential curve for three identical bosons is given
by
\begin{equation}  \label{U0bos}
U_0(R)=-\frac{s_0^2+1/4}{2\mu R^2}
\end{equation}
The coefficient $s_0\approx1.0062$ can be obtained analytically in the
theory of Efimov states~\cite{efim70,macek86,braaten2006ufb}. A simple and
fast numerical CGHS calculation with only 30 basis functions extended up to $%
R=100r_0$ shows, at large $R$, the expected $1/R^2$ behavior. Extrapolation
of our potential curves to $R\rightarrow\infty$ gives $s_0\approx1.0059$.

Similarly, we analyze the system of two indistinguishable fermions
resonantly interacting with a third particle of equal mass. For such system,
the zero-range model predicts a lowest potential of the form,
\begin{equation}  \label{U0ferm}
U_0(R)=\frac{p_0^2-1/4}{2\mu R^2}
\end{equation}
The value of $p_0\approx2.166222$ can also be predicted analytically. Using
a slightly larger basis set of 90 basis function we extend the CGHS
calculations up to $R=4000r_0$. Extrapolating our potential curves to $%
R\rightarrow\infty$ we obtain $p_0\approx2.166218$.

These two examples show that the CGHS method is flexible enough to describe
a strongly interacting system with relatively small basis sets and
analytical matrix element evaluations. The main limitation of these
calculations come from linear dependence issues. At the $N=3$ level, this
method cannot probably compete with more sophisticated calculations which
permit calculations up to $R=10^6r_0$~\cite{esry1999rta,suno2002tbr}.
However, it has been a challenge to extend hyperspherical methods beyond $%
N=3 $. One successful method uses Monte Carlo techniques to describe the
lowest channel function and extends it application to large ($N\lesssim10$)
systems~\cite{blume2000mch}. However, this method can only calculate the
lowest potential curve and leads to an approximate solution. In contrast,
the CGHS method can be readily extended to $N=4$ particles (and possibly
beyond) and allows to obtain a full solution which represents the current
\textit{state of the art} of hyperspherical methods.

\begin{figure}[htbp]
\begin{center}
\includegraphics[scale=0.6]{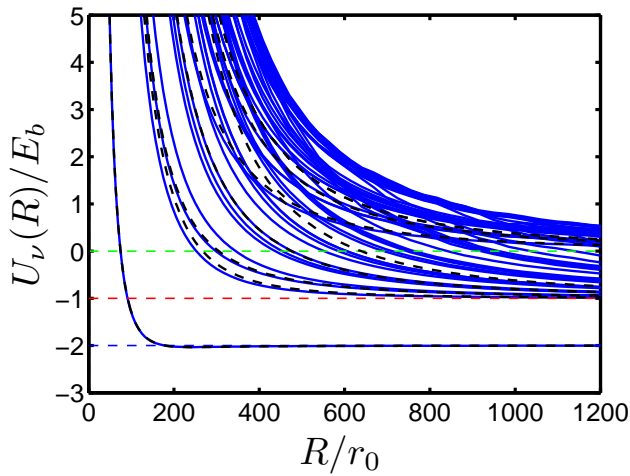}
\end{center}
\caption{(Color online) Adiabatic hyperspherical potential curves $U_\protect%
\protect\nu(R)$ (solid lines) for two spin-up and two spin-down fermions
with an atom-atom scattering length $a_s=100r_0$. The dashed line at $E=2E_b$
(blue online) is the dimer-dimer threshold, the dashed line at $E=E_b$ (red
online) is the dimer--two-atom threshold, and the dashed line at $E=0$
(green online) is the four-atom threshold. Dashed curves are predictions
from Ref.~\protect\cite{dincao08}.}
\label{PotCurvesBEC}
\end{figure}

The development of four-body hyperspherical methods allows, for one thing,
an analysis of the full energy dependence of the dimer-dimer scattering
length. Figure~\ref{PotCurvesBEC} presents the four-fermion potential curves
obtained with the correlated Gaussian hyperspherical method(CGHS).
 There are three relevant energy thresholds mark with dashed lines in Fig.~%
\ref{PotCurvesBEC}:dimer-dimer threshold at $2E_b$, dimer--two-atom
threshold at $E_b$ and four-atom threshold at 0 energy. The lowest curve
represents the dimer-dimer channel and potential curves going asymptotically
to $E_b$ and 0 represent dimer--two-atom and four-atom channels,
respectively. Standard multichannel scattering techniques, like the R-matrix method, can
be applied to solve the hyperspherical coupled differential equations. This
analysis was performed in a recent study by D'Incao \emph{et. al.}~\cite%
{dincao08}, which obtained the energy dependence of the dimer-dimer
scattering length for equal mass systems. Black dashed curves in Figure~\ref%
{PotCurvesBEC} represent the potential curves of Ref.~\cite{dincao08}. As we
can see, the CGHS method presented here predicts very similar potential
curves. The dimer-dimer potential curves obtained with the different methods
are almost indistinguishable. For dimer--two-atom potential curves, the CGHS
predicts lower potential curves suggesting that the CGHS calculation is
slightly better. At large $R$, the asymptotic behavior of both methods
agree. This is very encouraging since in the method of D'Incao \emph{et. al.}%
, the asymptotic behavior of the channel functions is correct by
construction, whereas in the CGHS it constitutes an important, nontrivial
test. Preliminary calculations with the CGHS potential curves predict a
similar energy dependence of the dimer-dimer scattering length. Therefore,
the CGHS opens the possibility for accurately analyzing four-body scattering
events, as has been carried out for four-interacting bosons in Refs.~\cite{vonstecher2008fbl,dincaoDimDim09}.

The calculations of the potential curves at unitarity allows us to extract
the four-fermion universal coefficients. As in the $N=3$ system, the
potential curves can be written as~\cite{tan05,wern06},
\begin{equation}
U_{\nu }(R)=\frac{p_{\nu }^{2}-1/4}{2\mu R^{2}}.  \label{U4ferm}
\end{equation}%
This functional form of the potential curves was verified indirectly in Ref.~%
\cite{blumePRL07} by
analyzing the spectrum of the four-fermion system under spherical harmonic
confinement. It can be shown that all the couplings vanish when the
potential curves are proportional to $1/R^{2}$. Therefore the system is
described by a set of uncoupled one-dimensional Schr\"{o}dinger equations
that can be solved analytically once the trapping potential is included.
These procedure leads to simple expressions for the trapped energies~\cite%
{wern06}
\begin{equation}
E_{\nu n}=(p_{\nu }+2n+5/2)\hbar \omega .  \label{E4ferm}
\end{equation}%
where $\omega $ is the trapping frequency and we have included the zero
point energy of the center-of-mass motion. In Ref.~\cite{blumePRL07}, the $%
2\hbar \omega $ spacing was verified and the lowest $p_{\nu }$ coefficients
were identified. Equations~(\ref{U4ferm}) and (\ref{E4ferm}) are also valid
in the non-interacting limit.
For $L=0$ and positive parity solutions, the $p_{\nu }^{NI}$ values and
their degeneracies $\lambda _{\nu }$ have relatively simple closed forms: $%
p_{\nu }^{NI}=11/2+2\nu $ and $\lambda _{\nu }=\nu ^{4}/96+7\nu
^{3}/48+17\nu ^{2}/24+133\nu /96+57/64+(-1)^{\nu }\nu /32+7/64(-1)^{\nu }$%
. Their lowest values can be found in Table~\ref{table0}.

\begin{table}[tbp]
\caption{Non interacting coefficients $p^{NI}_\protect\protect\nu$ of the
four-fermion potential curves and their degeneracies $\protect\lambda%
_\protect\protect\nu$. }
\label{table0}
\begin{center}
\begin{tabular}{||c|c|c||c|c|c||}
\hline
$\nu$ & $p^{NI}_{\nu}$ & $\lambda_\nu$ & $\nu$ & $p^{NI}_{\nu}$ & $%
\lambda_\nu$ \\ \hline
0 & 11/2 & 1 & 5 & 31/2 & 50 \\
1 & 15/2 & 3 & 6 & 35/2 & 80 \\
2 & 19/2 & 8 & 7 & 39/2 & 120 \\
3 & 23/2 & 16 & 8 & 43/2 & 175 \\
4 & 27/2 & 30 & 9 & 47/2 & 245 \\ \hline
\end{tabular}%
\end{center}
\end{table}

The development of the CGHS method allows us to carry out a hyperspherical
calculation for the four-fermion problem and to directly verify the form of
the hyperspherical potentials. Also, it allows us to analyze deviations from
the zero-range solutions due to finite-range effects.

\begin{figure}[htbp]
\begin{center}
\includegraphics[scale=0.6,angle=0]{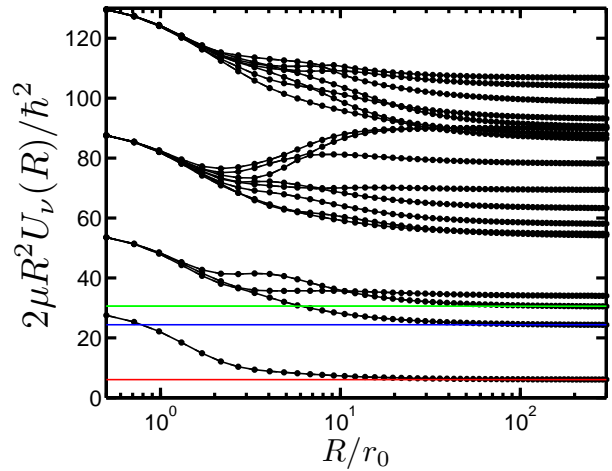}
\end{center}
\caption{(Color online) Hyperspherical potential curves at unitarity ($a\rightarrow \infty$%
) for the 4-fermion system multiplied by $2\protect\mu R^2/\hbar^2$. The
solid lines represent the predictions from analyzing the spectrum obtained
with the CG method. The symbols correspond to direct evaluation of the
potential curves with the CGHS method.}
\label{UnitPotCurves}
\end{figure}

The $20$ lowest four-body potential curves $2\mu R^2 U_\nu(R)/\hbar^2$ for
the equal-mass system are presented in Figure~\ref{UnitPotCurves}. We can
identify three regimes in these potential curves. The region $R\lesssim r_0$
is controlled by the kinetic energy. The kinetic energy effects are more
important than the interaction energy and the potential curves are well
approximated by the non-interacting potential curves. In other words, $2\mu
R^2 U_\nu(R)/\hbar^2\approx (p^{NI}_\nu)^2-1/4$ and the eigenchannels are
well approximated by the hyperspherical harmonics (see Sec.~\ref{hyperRepSec}%
). For that reason, there is a large degeneracy in the $R\lesssim r_0$
region which corresponds to the degeneracy of the $\Lambda^2$ operator.
Furthermore, the potential curves are, to a good approximation, proportional
to $1/R^2$. The second region is $r_0\lesssim R\lesssim 20 r_0$. In this
region both the kinetic and the interaction terms are important and finite
range effects are important. In the third region, $R\gtrsim 20r_0$, the
potential curves recover their universal behavior. The potential curves are,
again, approximately proportional to $1/R^2$. As $R/r_0$ increases,
finite-range effects tend to zero and we obtain the zero-range potential
curves at unitarity. Therefore, in this region, the eigenvalues of $2\mu R^2
U_\nu(R)/\hbar^2$ are approximately $(p^2_\nu-1/4)$. Thus, we can compare
these results with the ones deduced from trapped calculations for $%
r_0/a_{ho}=0.01$ presented in Ref.~\cite{blumePRL07}. The solid lines
correspond to $(p^2_0-1/4)$, $(p^2_1-1/4)$ and $(p_2^2-1/4)$ respectively~\cite{fnote3}.
 There is good agreement between the
predictions from the trapped system obtained with CG and the direct
computation of the potential curves through CGHS.

\begin{table}[tbp]
\caption{Coefficients $p_\protect\protect\nu$ of the four-fermion potential
curves. }
\label{table1}
\begin{center}
\begin{tabular}{||c|c||c|c||c|c||}
\hline
$\nu$ & $p_{\nu}$ & $\nu$ & $p_{\nu}$ & $\nu$ & $p_{\nu}$ \\ \hline
0 & 2.509 & 7 & 7.959 & 14 & 9.502 \\
1 & 4.944 & 8 & 8.341 & 15 & 9.648 \\
2 & 5.529 & 9 & 8.848 & 16 & 9.938 \\
3 & 5.846 & 10 & 9.292 & 17 & 10.205 \\
4 & 7.363 & 11 & 9.366 & 18 & 10.339 \\
5 & 7.402 & 12 & 9.5 & 19 & 10.482 \\
6 & 7.621 & 13 & 9.501 &  &  \\ \hline
\end{tabular}%
\end{center}
\end{table}

To quantify this last statement, we analyze the value of $p_0$. Several
groups~\cite{stech07, chan07,vonstechtbp,alhassid08,StetcuCM07} have tried
to benchmark the four-body value of $E_{00}$, which is simply related to $%
p_0 $. The calculations from Ref.~\cite{alhassid08} use zero-range
interactions explicitly and they report a value of $E_{00}\approx
(5.045\pm0.003)\hbar\omega$. To extract the $p_0$ value in the zero-range
limit, we carry out two different calculations. First, we study the $E_{00}$
energy obtained with the standard CG method as a function of the range of the
two-body interaction and then we extrapolate to zero-range limit. This
method was previously applied for the three-body system and the numerical
results agreed with the analytical predictions up to 7 digits~\cite%
{vonstech08a}. The same procedure applied to the four-body system, leads to $%
p_0\approx 2.5096$. The second calculation analyzes the long-range behavior
of the potential curves. To eliminate finite-range effects, we extrapolate
the potential curve $U_0(R)$ to $R/r_0\rightarrow\infty$. In this limit, $%
U_0(R)$ is characterized by a value $p_0\approx 2.5092$. These two different
methods provide a value of $p_0$ which agrees in four digits. These values
are slightly lower than $p_0\approx 2.545\pm0.003$ predicted in Ref.~\cite%
{alhassid08}. This suggests that the uncertainty in Ref.~\cite%
{alhassid08} was apparently underestimated.

The calculations of the lowest 20 universal coefficients $p_\nu$ is reported
in Table~\ref{table1}. It is interesting to note that some of the $p_\nu$
coefficients are very similar to the noninteracting coefficients. For
example, the $p_\nu$ coefficient for $\nu$=12, 13, 14 coincide with
noninteracting $p^{NI}_\nu$ coefficient. Two of these potential curves are
also described by $p_\nu=9.5$ in the small $R$ region and deviate from these
value in the region $R\sim r_0$. These channels have nodes in every
spin-up--spin-down interparticle distance, therefore at large distances they
recover the noninteracting behavior. The third potential curve smoothly
decrease from $p^{NI}_\nu=11.5$ at small $R$ to $p_\nu=9.5$ at large $R$.

Finally, note that the CGHS method has been successfully applied to the
four-boson system~\cite{vonstecher2008fbl}. In that study, the four-boson
spectrum is calculated from the CGHS potential curves. Also, that study
considers scattering events such as four-body recombination, which was
calculated and predicted to be important for the understanding of a recent
experiment on Efimov physics in an ultracold Bose gas~\cite{kraemer2006eeq}.

\section{Conclusions}

\label{conclusions}

We have presented an innovative numerical method suitable for the analysis
of four-body processes. We have shown several examples for three- and
four-particle systems recovering known results. Furthermore, we have
obtained the lowest 20 $p_\nu$ coefficients for the two-component
four-fermion system that characterize both free and trapped systems. These
coefficients also characterize the spectrum of four trapped fermions at
unitarity. Our results considerably extend previous calculations and provide
more accurate energies.

The CGHS method has been used to analyze the four-boson system in
Refs.~\cite{vonstecher2008fbl,dincaoDimDim09,mehtaNbody09},
predicting new phenomena observed experimentally~\cite{grim4b09}.
 It has also built a theoretical foundation for the
analysis of four-body collisional processes in other systems such as
two-component Bose systems~\cite%
{papp:040402,myatt1997pto,thalhammer:210402,weber2008aud}, Bose-Fermi
mixtures~\cite{modugno2002cdf,inouye2004ohf,olsen2008cam}, and
three-component Fermi gases\cite{ottenstein:203202,huckans2008eri,
blume2008sim}. Even though this method was initially implemented to treat
ultracold systems using model potentials, it can be in principle extended to
other four-body problems.

The authors would like to thank S. T. Rittenhouse,
N. P. Mehta and J. P. D'Incao for useful discussions and for providing their
four-fermion numerical data (dashed curves included in Fig.~\ref%
{PotCurvesBEC}).  This work was supported in part by NSF.

\appendix

\section{Application of Correlated Gaussians to the hyperspherical framework}

\label{App} This appendix illustrates how correlated Gaussian basis
functions can be used in the general hyperspherical framework presented in
Subsec.~\ref{Exp}. First, we consider the three-particle case and calculate
the matrix elements (Eqs.~\ref{Hrr}, \ref{Orr} and \ref{PQrr}). Then, we
discuss how to generate and optimize the basis set.

\subsection{Unsymmetrized matrix element evaluation for three particles}

\label{unsym3}

In this subsection, we present as an example the evaluation of the matrix
elements (Eqs.~\ref{Hrr}, \ref{Orr}, \ref{PQrr}) of three particle system.
Consider a system in which the center-of-mass motion decouples. Then, the $%
L^P=0^+$ solutions of the body-fixed system can be expanded in terms of the
interparticle distances. For the three-body system the correlated basis
functions take the form
\begin{equation}  \label{Basis1I}
\Psi_A(r_{12},r_{13},r_{23})=\exp\left[-\left(\frac{r_{12}^2}{2d_{12}^2}+%
\frac{r_{13}^2}{2d_{13}^2}+\frac{r_{23}^2}{2d_{23}^2}\right)\right].
\end{equation}
For equal mass systems, we can write Eq.(\ref{Basis1I}) in terms of the
following Jacobi coordinates:
\begin{gather}  \label{Jaccor}
\mathbf{x_1}=\frac{1}{\sqrt{2}}(\mathbf{r}_1-\mathbf{r}_2), \\
\mathbf{x_2}=\sqrt{\frac{2}{3}}\left(\mathbf{r}_3-\frac{\mathbf{r}_1+\mathbf{%
r}_2}{2}\right).
\end{gather}
The basis functions [Eq.~(\ref{Basis1I})] can be written as
\begin{multline}  \label{Basis}
\Psi_A(r_{12},r_{13},r_{23})=\braket{\mathbf{x_1},\mathbf{x_2}|A}=\exp(-%
\frac{\mathbf{x}^T.A.\mathbf{x}}{2}) \\
=\exp(-\frac{\mathbf{x_1}.\mathbf{x_1}a_{11}+2\mathbf{x_1}.\mathbf{x_2}%
a_{12} +\mathbf{x_2}.\mathbf{x_2} a_{22}}{2})
\end{multline}
where $\mathbf{x}\equiv\{\mathbf{x}_1,\mathbf{x}_2\}$ and $A$ is a 2 by 2
symmetric matrix whose elements are $%
a_{11}=2/d_{12}^2+1/2(1/d_{13}^2+1/d_{23}^2)$ , $a_{12}=a_{21}= \sqrt{3}/2
(1/d_{23}^2-1/d_{13}^2)$, and $a_{22}=3/2(1/d_{13}^2 + 1/d_{23}^2)$. In Eq.~(%
\ref{Basis}), we can clearly see that the state $\braket{\mathbf{x_1},%
\mathbf{x_2}|A}$ depends only on the distances $x_1$, and $x_2$ plus the
angle $\theta_{12}$ between them, $\cos\theta_{12}=\mathbf{x_1}.\mathbf{x_2}%
/x_1 x_2$.

We want to obtain the matrix elements corresponding to these basis functions
at fixed hyperradius $R$. We define the hyperradius to be $R^2=\mathbf{x}%
_1^2+\mathbf{x}_2^2$. The integrand of the overlap matrix element between $%
\ket{A}$ and $\ket{B}$, noted as $B.A$, is
\begin{equation}  \label{O1}
B.A=\exp(-\frac{\mathbf{x}^T.(A+B).\mathbf{x}}{2}).
\end{equation}
We change to the Jacobi basis set that diagonalizes $A+B$ and we call $%
\beta_1$ and $\beta_2$ the eigenvalues and $\mathbf{y}\equiv\{\mathbf{y}_1,%
\mathbf{y}_2\}$ the orthonormal eigenvectors. In this new coordinate basis,
Eq.~(\ref{O1}) has a simple form,
\begin{equation}  \label{O2}
B.A=\exp(-\frac{\beta_1y_1^2+\beta_2y_2^2}{2}).
\end{equation}
We integrate over the angles of the vectors $\mathbf{y}_1$ and $\mathbf{y}_2$
and we fix the hyperradius, so $y_1=R\cos\theta$ and $y_2=R\sin\theta$. In
this set of coordinates, the matrix element at fixed $R$ is
\begin{equation}  \label{O3}
\braket{B|A}\Big|_R=(4
\pi)^2\int_0^{\pi/2}e^{-R^2(\beta_1\cos^2\theta+\beta_2\sin^2\theta)/2}%
\cos^2\theta\sin^2\theta d\theta
\end{equation}
This integration has a closed-form result,
\begin{equation}  \label{O4}
\braket{B|A}\Big|_R=2 \pi^3\frac{\exp(-\frac{\beta_1+\beta_2}{4} R^2)}{\xi}
I_1\left(\xi\right) .
\end{equation}
Here we have introduced the definition $\xi=R^2(\beta_1-\beta_2)/4$.

To simplify the interaction matrix element evaluation, we can adopt a
Gaussian model potential as was utilized in the CG method. In this case, the
interaction term can be evaluated in the same way we have calculated the
overlap term since the interaction is also a Gaussian. Each pairwise
interaction can be easily written as $V_{ij}=V_0\exp(-\frac{r_{ij}^2}{2r_0^2}%
)=V_0 \exp(-\mathbf{x}^T.M^{(ij)}.\mathbf{x}/(2r_0^2))$. Therefore, to
calculate the interaction matrix element, we need to evaluate
\begin{equation}  \label{V1}
\braket{B|V_{ij}|A}=V_0\int d\Omega \exp(-\frac{\mathbf{x}%
^T.(A+B+M^{(ij)}/r_0^2).\mathbf{x}}{2}).
\end{equation}
This integration can be done following the same steps of the overlap matrix
element. Equation~(\ref{O4}) can be used directly if we multiply it by $V_0$%
, and $\beta_1$ and $\beta_2$ are replaced by the eigenvalues of $%
A+B+M^{(ij)}/r_0^2$. Note that for each pairwise interaction (and for each
pair of basis functions in the matrix element), the matrix $M^{(ij)}$
changes and requires a new evaluation of the eigenvalues.

The third term we need to evaluate is the hyperangular kinetic term at fixed
$R$. This kinetic term is proportional to the grand angular momentum
operator $\Lambda$ defined for the $N=3$ case as
\begin{equation}
\frac{\Lambda^2\hbar^2}{2\mu R^2}=-\sum_i\frac{\hbar^2\nabla_i^2}{2\mu}+%
\frac{\hbar^2}{2\mu}\frac{1}{R^{5}} \frac{\partial}{\partial R} R^{5}\frac{%
\partial}{ \partial R}.
\end{equation}
The expression can be formally written as
\begin{equation}  \label{To}
\mathcal{T}_\Omega=\mathcal{T}_T-\mathcal{T}_R,
\end{equation}
where
\begin{equation}  \label{Tto}
\mathcal{T}_\Omega=\frac{\Lambda^2\hbar^2}{2\mu R^2},\;\;\;\;\mathcal{T}%
_T=-\sum_i\frac{\hbar^2\nabla_i^2}{2 \mu},
\end{equation}
and
\begin{equation}  \label{Tto2}
\mathcal{T}_R=-\frac{\hbar^2}{2\mu}\frac{1}{R^{5}} \frac{\partial}{ \partial
R} R^{5}\frac{\partial}{ \partial R}.
\end{equation}

In typical calculations, $\mathcal{T}_\Omega$ is evaluated by directly
applying the corresponding derivatives in the hyperangles $\Omega$. However,
in this case, it is convenient to evaluate $\mathcal{T}_T$ and $\mathcal{T}%
_R $ separately, and make use of~(\ref{To}).

The integrand of the total kinetic term $\mathcal{T}_T$ takes the form
\begin{equation}  \label{TtE1}
B|\mathcal{T}_T| A=\exp(-\frac{\mathbf{x}^T.B.\mathbf{x}}{2})\left(-\sum_i^2%
\frac{\hbar^2}{2\mu}\nabla^2_i\right)\exp(-\frac{\mathbf{x}^T.A.\mathbf{x}}{2%
}).
\end{equation}

First, we diagonalize $A$ and use the eigenvectors and eigenvalues of $A$,
obtaining,
\begin{equation}
B|\mathcal{T}_T| A=-\frac{\hbar^2}{2\mu} \left(-\mbox{Tr}[A]+ \mathbf{x}%
^T.A^2.\mathbf{x}\right)\exp(-\frac{\mathbf{x}^T.(A+B).\mathbf{x}}{2}).
\end{equation}
Here $\mbox{Tr}$ is the trace function. We can use $\mbox{Tr}%
[A]=(\alpha_1+\alpha_2)$, where $\alpha_1$ and $\alpha_2$ are the
eigenvalues of $A$. Now we diagonalize $A+B$. We call $T$ the matrix with
the orthonormal eigenstates in columns and $\beta_1$ and $\beta_2$ are the
eigenvalues of $A+B$. We make a change of coordinates to the basis set that
diagonalizes $A+B$. We obtain
\begin{equation}
B|\mathcal{T}_T| A=-\frac{\hbar^2}{2\mu} \left(-3(\alpha_1+\alpha_2)+
\mathbf{y}.G.\mathbf{y}\right)\exp(-\frac{\beta_1 y_1^2+\beta_2 y_2^2}{2}),
\end{equation}
where $G=T^T.A^2.T$, and $\mathbf{y}_1$ and $\mathbf{y}_2$ are the vectors
in the new eigen basis. The integration over the angles of these vectors is
trivial. After this integration, we fix the hyperradius and integrate over
the hyperangle $\theta$ defined by $y_1=R\cos\theta$ and $y_2=R\sin\theta$,
\begin{multline}
\braket{B|\mathcal{T}_T| A}\Big|_R=-\frac{(4 \pi)^2\hbar^2}{2\mu}%
\int_0^{\pi/2} \Big[-3(\alpha_1+\alpha_2) \\
+ g_{11}R^2 \cos^2\theta+ g_{22}R^2 \sin^2\theta\Big] \\
\exp(-\frac{\beta_1 R^2 \cos^2\theta+\beta_2 R^2 \sin^2\theta}{2})
\cos^2\theta\sin^2\theta d\theta.
\end{multline}
This integration can be done analytically and the results expressed in terms
of the Bessel functions $I_1$ and $I_0$:
\begin{multline}  \label{TtE3}
\braket{B|\mathcal{T}_T| A}\Big|_R=-\frac{\hbar^2e^{-\frac{%
(\beta_1+\beta_2)R^2}{2}}\pi^3R^2}{16 \xi\mu} \\
\left\{-8(g_{11}-g_{22}) I_0\left[\xi\right] +\frac{2}{\xi}\Big\{%
8(g_{11}-g_{22})\right. \\
\left.+(\beta_1-\beta_2)\left(-6(\alpha_1+\alpha_2)+(g_{11}+g_{22})R^2\right)%
\Big\}I_1\left[\xi\right]\right\}.
\end{multline}

Now we will evaluate $\mathcal{T}_R$, the hyperradial kinetic term. It is
written as
\begin{equation}  \label{Tr}
\mathcal{T}_R=-\frac{\hbar^2}{2\mu} \left(\frac{1}{R^{5/2}} \frac{\partial^2%
}{\partial R^2}R^{5/2}- \frac{15}{4 R^2}\right).
\end{equation}

Therefore, the integrand takes the form
\begin{multline}  \label{TrE1}
B|\mathcal{T}_R| A=-\frac{\hbar^2}{2\mu}\exp(-\frac{\mathbf{x}^T.B.\mathbf{x}%
}{2}) \\
\left(\frac{1}{R^{5/2}} \frac{\partial^2}{\partial R^2}R^{5/2}- \frac{15}{4
R^2}\right)\exp\left(-\frac{\mathbf{x}^T.A.\mathbf{x}}{2}\right),
\end{multline}

We use the property $\mathbf{x}^T.A.\mathbf{x}=R^2 F_A(\Omega)$ to evaluate
the derivatives with respect to $R$. This allows a simple calculation of the
derivatives in Eq.~(\ref{TrE1}), yielding
\begin{equation}  \label{TrE2}
B|\mathcal{T}_R|A=\frac{\hbar^2}{2\mu R^2} \left[6 \mathbf{x}^T.A.\mathbf{x}%
-(\mathbf{x}^T.A.\mathbf{x})^2 \right] e^{-\mathbf{x}^T.(A+B).\mathbf{x}/2}.
\end{equation}
Next we diagonalize $A+B$, and set $D=T^T.A.T$, giving
\begin{equation}
B|\mathcal{T}_R|A=\frac{\hbar^2}{2\mu R^2} \left[6 \mathbf{y}.D.\mathbf{y}-(%
\mathbf{y}.D.\mathbf{y})^2 \right]\exp\left(-\frac{\beta_1 y_1^2+\beta_2y_2^2%
}{2}\right)
\end{equation}
The terms $\mathbf{y}.D.\mathbf{y}$ and $(\mathbf{y}.D.\mathbf{y})^2$ depend
on the polar angles of the vectors. The integration over the polar angles ($%
\Omega_1=\{\phi_1,\theta_1\}$ and $\Omega_2=\{\phi_2,\theta_2\}$)
of these terms is
\begin{multline}  \label{TrE4}
\int\left[6 \mathbf{y}.D.\mathbf{y}-(\mathbf{y}.D.\mathbf{y})^2 \right]
d\Omega_1d\Omega_2= (4\pi)^2\Big\{6d_{11}y_1^2+6d_{22}y_2^2 \\
-\big[d_{11}^2 y_1^4+(2d_{11}d_{22}+4d_{12}^2/3)y_1^2y_2^2+d_{22}^2 y_2^4%
\big]\Big\}.
\end{multline}
Now we carry out the integration over the hyperangle $\theta$, using $%
y_1=R\cos\theta$ and $y_2=R\sin\theta$, which gives
\begin{multline}
\braket{B|\mathcal{T}_R|A}\Big|_R=\frac{(4\pi)^2\hbar^2}{2\mu R^2}%
\int_0^{\pi/2} \Big\{6d_{11}R^2\cos^2\theta \\
+6d_{22}R^2\sin^2\theta-d_{11}^2 R^4\cos^4\theta \\
-(2d_{11}d_{22}+4d_{12}^2/3)R^4\cos^2\theta\sin^2\theta-d_{22}^2 R^4
\sin^4\theta\Big\} \\
\exp\left(-\frac{\beta_1R^2\cos^2\theta+\beta_2R^2\sin^2\theta}{2}\right)
\cos^2\theta\sin^2\theta d\theta.
\end{multline}

This integration has the analytical form

\begin{multline}  \label{TrE3}
\braket{B|\mathcal{T}_R|A}\Big|_R= -\frac{\hbar^2}{\mu}\frac{e^{-\frac{%
(\beta_1+\beta_2) R^2}{4}}\pi^3 R^2 }{64 \xi^2} \Big\{-8\Big[-8 d_{12}^2 \\
+(d_{11}-d_{22})\big(6(-\beta_1+\beta_2+d_{11}-d_{22})+ \\
4\xi(d_{11}+d_{22})\big)\Big]I_0[\xi]+ \frac{2}{\xi}\Big[-64d_{12}^2+ \\
48(d_{11}-d_{22})(-\beta_1+\beta_2+d_{11}-d_{22}) \\
+8\xi(-3\beta_1+3\beta_2+4d_{11}-4d_{22})(d_{11}+d_{22}) \\
\left.+16\xi^2(d_{11}^2+d_{22}^2)\Big]I_1[\xi]\right\}.
\end{multline}

Combining Eqs.~(\ref{TtE3}, \ref{TrE3}), we obtain $\mathcal{T}_\Omega$. The
expression for $\mathcal{T}_\Omega$ can be simplified using the relation $%
G=D^2$ to write $G$ matrix elements of Eq.~(\ref{TtE3}) in terms of the ones
of $D$. This same procedure can be applied to extract the $P$ and $\tilde{Q}$ matrix
elements:
\begin{equation}  \label{HRa2}
\braket{B|\frac{\partial A}{\partial R}}\Big|_R, \,\,\,\mbox{and}\,\,\,\,\braket{\frac{\partial
B}{\partial R}|\frac{\partial A}{\partial R}}\Big|_R.
\end{equation}

A useful test to verify the functional form of the matrix elements is to
integrate them with respect to $R$, with the corresponding volume element,
and compare that result with the standard CG matrix elements. Another
important test is to verify that $\mathcal{T}_\Omega$ is symmetric under the
exchange of the basis functions $A$ and $B$. This is not a trivial test
since neither $\mathcal{T}_T$ nor $\mathcal{T}_R$ are symmetric.

A major advantage of these matrix element evaluations is that they can be
easily extended to four particles. In general, these matrix elements
evaluations would require a 5-D numerical integration but for these basis
functions, with the above analytical development, they only require a 1-D
numerical integration.

\subsection{General considerations}

\label{GenCon}

Many of the procedures of the standard CG method can be easily extended to
the CGHS. The selection, symmetrization, and optimization of a basis follow
the same ideas of the standard CG method. However, the evaluation of the
unsymmetrized matrix elements at fixed $R$ is clearly different.
Furthermore, the hyperangular Hamiltonian [Eq.~\ref{LamHRnew}] need to
solved at different hyperradius $R$.

There are several properties that make this method particularly efficient.
For the model potential used, the scattering length is tuned by varying the
potential depths of the two-body interaction. Therefore, as in the CG case,
the matrix elements need only be calculated once; then they can be used for
a wide range of scattering lengths. Of course, the basis set should be
complete enough to describe the relevant potential curves at all the desired
scattering length values.

The selection of the basis function generally depends on $R$. To avoid
numerical problems, the mean hyperradius of each basis function $\braket{R}%
_B $ should be comparable to the hyperradius $R$ in which the matrix
elements are evaluated. We can ensure that $\braket{R}_B\sim R$ by selecting
some (or all) the weights $d_{ij}$ to be of the order of $R$.

We consider two different optimization procedures. The first possible
optimization procedure is the following: First, we select a few basis
functions and optimize them to describe the lowest hyperspherical harmonics.
The Gaussian widths of these basis functions are rescaled by $R$ at each
hyperradius so that they represent the hyperspherical harmonics equally
well. These basis functions are used at all $R$, while the remaining are
optimized at each $R$. Starting from small $R$ (of the order of the range of
the potential), we optimize a set of basis functions. As $R$ is increased,
the basis set is increased and reoptimized. At every $R$ step, only a
fraction of the basis set is optimized, and those basis functions are
selected randomly. After a several $R$-steps, the basis set is increased.

Instead of optimizing the basis set at each $R$, one can alternatively try
to create a complete basis set at large $R_{max}$. In this case, the basis
functions should be complete enough to describe the lowest channel functions
with interparticle distances varying from interaction range $r_0$ up to the
hyperradius $R_{max}$. Such a basis set can be rescaled to any $R<R_{max}$
and should efficiently describe the channel functions at that $R$. The
rescaling procedure is simply $d_{ij}/R=d_{ij}^{max}/R_{max}$. This
procedure avoids the optimization at each $R$. Furthermore, the kinetic,
overlap, and couplings matrix elements at $R$ are straightforwardly related
with the ones at $R_{max}$. So, the interaction potential is the only matrix
element that needs to be recalculated at each $R$. This property can be
understood using dimensional analysis. The kinetic, overlap, and coupling
matrix elements only depend on $R$, so a rescaling of the widths is simply
related to a rescaling of the matrix elements. In contrast, the interaction
potential introduces a new length scale, so the matrix elements depend on
both $R$ and $r_0$, and the rescaling does not work.

These two methods, the ``complete basis set'' or the ``small optimized basis
set'' method, can be appropriate in different circumstances. If a large
number of channels are needed, probably the complete basis method is the
best choice. But, if only a couple of particular channel potential curves
and couplings are needed, then the small optimized basis set method might be
more efficient.

The most convenient strategy we have found for optimizing the basis function
in the four-boson and four-fermion problems is the following: First we
select an hyperradius $R_m$ that is $R_m\approx 300\, r_0$ where the basis
function will be initially optimized. The basis set is increased and
optimized until the relevant potential curves are converged and, in that
sense, the basis is complete. This basis is then rescaled, as proposed in
the second optimization method, to all $R<R_m$. For $R>R_m$, it is too
expensive to have a ``complete" basis set. For that reason, we use the
``small optimized basis set'' method which allows a reliable description of
the lowest potential curves.

Note that for standard correlated Gaussian calculations, the matrices $A$
and $B$ need to be positive definite. This condition restricts the Hilbert
space to exponentially decaying functions. In the hyperspherical treatment,
this is not necessary since the matrix elements can always be calculated at
fixed $R$, as the integrals converge even for exponentially growing
functions. This gives more flexibility in choosing the optimal basis
functions.


\begin{thebibliography}{10}
\bibitem{fnote1}
Even though $\mu$ is arbitrary, there is one preferable selection:
  $\mu=[\prod_i m_i/(\sum_i m_i)]^{1/(N-1)}$ that conserves the volume element
  of the coordinate transformation.\cite{Delves60} In this article, we will
  only present results for equal mass systems for which is simpler to simply
  choose $\mu=m$.
\bibitem{fnote2}
The (3N-7) numerical integration results from the following reasoning:
  initially we have 3N numerical integration but 3 dimensions are removed by
  decoupling the center of mass motion, 3 dimension are removed fixing the
  Euler angles and 1 dimension is removed by fixing R.
\bibitem{fnote3}
The coefficients $p_\nu$ can be easily related with the $s_\nu$ of
  Ref.~\cite{blumePRL07} by $s_\nu=p_{\nu-1}-1/2$.
\end{thebibliography}

\end{document}